\documentclass[10pt,letterpaper,twocolumn]{article} 

\usepackage{ol2}
\usepackage[draft]{hyperref}
\usepackage{amsmath}

\begin{document}
\twocolumn[

\title{Fiber based source of photon pairs at telecom
band with high temporal coherence and brightness for quantum
information processing}

\author{Xiaoying Li$^{\text{1}*}$, Lei Yang$^{\text{1}}$, Liang Cui$^{\text{1}}$, Zhe Yu Ou$^{\text{2}\ddagger} $ and Daoyin Yu$^{\text{1}}$}

\address{$^{\text{1}}$ College of Precision Instrument and
Opto-electronics Engineering, Tianjin University, Tianjin, 300072,
P. R. China}
\address{$^{\text{2}}$ Department of Physics, Indiana University-Purdue University Indianapolis, Indianapolis, IN, 46202, USA}


\begin{abstract}We experimentally demonstrate a bright pulsed source of
correlated photon pairs at 1550 nm telecom band by pumping 300\,m
dispersion shifted fiber with a 4 ps pulse train. We investigate
the coherence property of the source by measuring the second order
intensity correlation function $g^{(2)}$ of individual signal
(idler) photons. A preliminary Hong-Ou-Mandel type two-photon
interference experiment with two such sources confirms the high
temporal and spatial coherence of the source. The source is
suitable for multi-photon quantum interference of independent
sources, required in quantum information
processing.
\end{abstract}


] 

\noindent Quantum interference (QI) among independent sources of entangled
photon pairs plays an important role in quantum information
processing (QIP), such as entanglement swapping, quantum teleportation,
generation of GHZ states, quantum repeater and linear optical
quantum computing ~\cite{Bouwmeester97,Pan98,Knill01,Zhao04}.
Because of the independent nature of the photons, in order to
obtain high visibility in QI, photons must be in identical single
mode so that they are indistinguishable. Nowadays, the most
popular sources of photons are realized via spontaneous parametric
down-conversion (SPDC) by pumping a $\chi^{(2)}$ nonlinear crystal
with ultrashort pulses. Although the SPDC parametric processes are
independent, the timing provided by the ultrashort pump pulse can
be used to realize synchronization of different sources of
independent photons~\cite{Zukowski95}. However, the dispersive
nature of the $\chi^{(2)}$ crystal completely messes up the
temporal mode structure of the down-converted fields. To preserve
the the temporal indistinguishability, optical filtering is
necessary so that the coherence time of the down converted signal
(idler) photons is much longer than the pump pulse duration. This
requirement, coupled with the low SPDC efficiency in $\chi^{(2)}$
crystal, severely limits the brightness of the photon pairs. It
should be mentioned that efforts are made recently to engineer the
structure of the $\chi^{(2)}$ materials to tailor their temporal
mode structure~\cite{Mosley07}.

Recently there are growing interests in generating photon pairs in optical fiber by means of four wave
mixing (FWM)
~\cite{Li05a,Lee06,Takesue06,Fan05a,Fulconis05}, because of
its inherent compatibility with the transmission medium, the
excellent single spatial-mode purity, and a better nonlinear vs.
loss figure of merit over $\chi^{(2)}$ nonlinear crystal. Because
of the near degenerate nature of FWM in dispersion shifted fiber (DSF),
when the pump wavelength is close to the zero dispersion wavelength $\lambda
_0$, dispersion plays a relatively small role in phase
matching in the sense that the band width of the signal and idler
fields are exceptionally wide even for a fiber with a length of more than hundreds meters~\cite{Li07}.
This leads to very simple spectral structure and
relatively mild filtering is enough to achieve good temporal
coherence. Indeed, recent Hong-Ou-Mandel (HOM) type experiments
using two pairs of visible photons originated from independent
MF-based sources~\cite{Fulconis07} was reported with a visibility of
95\%. However, for the photons at 1550nm telecom band produced by
two independent DSF-based sources~\cite{takesuehom}, a visibility
of only about 20\% and 50\% was obtained in a two-fold and
four-fold coincidence measurement, respectively. The result was
far from the ideal situation achievable by using photons with a
good spatial and temporal mode.

In this paper, we study the coherence property of light at telecom band generated in
DSF by measuring the normalized intensity
correlation function $g^{(2)}$ under various conditions for
individual signal (idler) field. According to
Refs.\cite{Ou99} and \cite{ou07}, $g^{(2)}$ of individual
signal and idler fields is related directly to the coherence
property of the field. We demonstrate that the source has high
temporal coherence and brightness. Moreover, to illustrate its
potential application for QIP, a two-photon HOM type interference experiment is performed with signal photons
originated from two DSF-based independent sources. A visibility of $33\%$ in
the HOM interference dip is observed, consistent with the thermal
nature of the individual signal (idler) field in single mode.

\begin{figure}[htb]
\includegraphics[width=8cm]{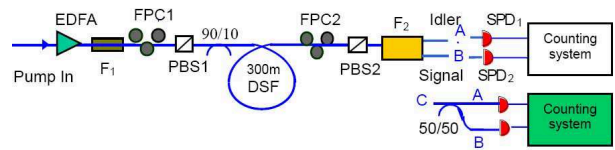}
\caption{(Color online) A schematic for characterizing our fiber
based source.} \label{setup-fig1}
\end{figure}

A schematic to characterize our fiber based source is shown in
Fig.~\ref{setup-fig1}. The pump pulses with a pulse width of $\sim 4$\,ps are spectrally
carved out from a mode-locked femto-second fiber laser with a
repetition rate of 40 MHz. To achieve the required power, the pump
pulses are amplified by an erbium-doped fiber amplifier.
The pump pulses are further cleaned up with a band-pass filter
F$_1$, which has various bandwidth by cascading a combination of
tunable filters (TF) (Newport/TBF-1550-1.0, Santec/OTM-30M-03D)
and one channel of an array-waveguide-grating (AWG). The pump is then passed through a fiber
polarization controller (FPC1) and a polarization beam splitter
(PBS1) to ensure its polarization and power adjustment. A 90/10 fiber coupler is used
to split $10\%$ of the pump for power monitor.

Signal and idler photons at 1546.9\,nm and 1530.9\,nm are produced
by pumping 300\,m DSF with laser pulses having a central wavelength $\lambda _p=
1538.9$\,nm. The DSF with $\lambda
_0=1538\pm 2$\,nm is submerged in liquid nitrogen (77\,K) to
reduce the Raman scattering (RS). Signal and idler photons co-polarized with the pump are selected by adjusting FPC2 placed in front of PBS2.
To reliably detect the signal and idler photons, an
isolation between the pump and signal/idler photons in excess of
100\,dB is required, because of the low efficiency of spontaneous
FWM in DSF. In addition, detecting signal (idler) photons with
different bandwidth is also necessary to characterize the source.
We achieve these by passing the photon pairs through a filter
ensemble F$_2$ which is realized either by using double grating
filters (DGFs)~\cite{Li05a} or TF, or by cascading a DGF and one channel of AWG.

The signal (idler) photons are counted by single photon
detectors (SPD, id200) operated in the gated Geiger mode. The $2.5$\,ns
gate pulses arrive at a rate of about 600 KHz, which is $1/64$ of the
repetition rate of the pump pulses, and the dead time of the gate
is set to be 10 $\mu$s.

\begin{figure}[htb]
\includegraphics[width=8cm]{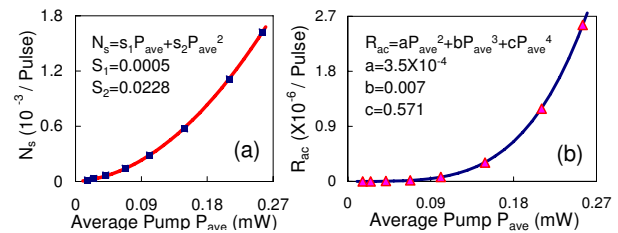}
\caption{(Color online) (a) The number of scattered photons per
pump pulse detected in the signal channel, $N_{s}$,  versus pump power.(b)
Accidental-coincidence rate R$_{ac}$ as a function of the pump power. The
solid lines are fit to corresponding polynomials.}
\label{results-fig2}
\end{figure}

We first test the brightness of the source by directly sending the
signal and idler to A and B channels, respectively. The full-width-half-maximum (FWHM) of pump is about 0.9\, nm, set by F$_1$ with two cascading TFs. The filter
ensemble $F_2$ consists of the AWG and DGF with a combined
FWHM of about $0.33$\,nm. We measure the number of scattered
photons in signal (idler) band per pump pulse, $N_{s(i)}$ as a function of the
average pump power, $P_{ave}$ [Fig.~\ref{results-fig2}(a)] and
fit the measured data with $N_{s(i)}=s_1P_{ave}+s_2P_{ave}^2$,
where $s_1$ and $s_2$ are the linear and quadratic coefficients,
which respectively determine the strengths of RS and FWM in DSF.
The fitting result [Fig.~\ref{results-fig2}(a)] shows that the number of photons via RS is much
less than that via FWM. To demonstrate that the background of our
source is low enough to allow a multi-photon pair experiment, we
measure the accidental coincidence between the signal and idler
fields by put a delay between A and B so that A and B register
photons from adjacent pulses. We plot the accidental coincidence
rate R$_{ac}$ as a function of the average pump power
[Fig.~\ref{results-fig2}(b)] and fit the experimental data to a
polynomial function $R_{ac}=aP_{ave}^2+bP_{ave}^3+cP_{ave}^4$,
where the coefficients $a$, $b$ and $c$ are the weights of the
coincidence contributed by the random overlap between Raman
photons and Raman photons, Raman photons and FWM photons, and FWM
photons and FWM photons, respectively.  The result of the fit
clearly shows that R$_{ac}$ is dominated by the event of
multi-photon pairs originated from FWM. Therefore, the coincidence
rate presented here is an indication of double pair or four-photon
coincidence rate, which is rather high at only a pump power of 0.25\,mW.

Next, we measure the photon bunching effect or $g^{(2)}$ of
individual signal (idler) photons. The photon bunching effect in
spontaneous parametric process was proven to be closely associated
with the visibility of QI involving
independent photon pairs~\cite{Ou99,Rarity97,Rarity98} because of
its connection to the coherence property of the fields
\cite{Ou99,ou07}. In the experiment, one output of our fiber
source, i.e., signal (idler) photons, is feeded to the port
labeled "C" and passes though a 50/50 beam splitter (BS) (see Fig.
1). The two outputs of the BS are detected by SPD1 and SPD2,
respectively, and both the coincidence and accidental coincidence
of the two SPDs are recorded. The second order correlation
function $g^{(2)}$ is the ratio between the measured coincidence
and accidental-coincidence rates. The result shows that when the
bandwidth of the pump and the detected field are 0.9\,nm and 0.33\,nm,
respectively, the value of $g^{(2)}$ for the individual signal
(idler) photons is $1.97\pm 0.03$, which is very close to that of
single mode field, $g^{(2)}=2$, indicating the signal (idler)
photons produced by the fiber source can be viewed as a single
spatial and temporal mode~\cite{Ou99}.

\begin{figure}[htb]
\includegraphics[width=7cm]{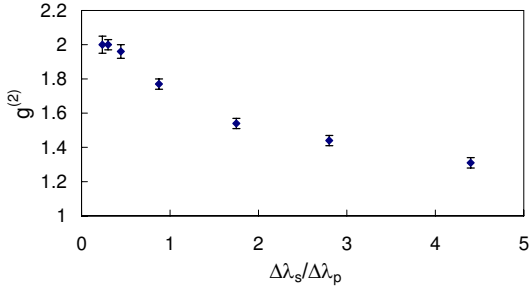}
\caption{(Color online) The second correlation function $g^{(2)}$
versus the ratio between the bandwidth of the signal (idler)
photons and that of the pump photons. The FWHM of signal and pump photons is denoted by $\Delta \lambda_{s}$ and $\Delta \lambda_p$, respectively.} \label{3}
\end{figure}

Although setting the central wavelength of pump close to $\lambda
_0$ of DSF makes phase matching a less
important factor, filters are still necessary to clean up the
temporal modes~\cite{Li07}. We next investigate the dependence of $g^{(2)}$
on the bandwidth of signal (idler) and pump photons. Using various
filters ranging from $0.26$\,nm to $1.3$\,nm for the pump and
signal (idler) fields, we measure $g^{(2)}$ when the peak power of
the pump is about $1$\,W. As shown in Fig.\ref{3}, the value of
$g^{(2)}$ depends highly on the ratio between the optical
bandwidths of photon pairs and that of pump pulses: $g^{(2)}$
approaches to $2$ when the ratio is small and drops to close $1$
as the ratio increases. This is consistent with a simple theory on
the coherence of the individual signal and idler fields~\cite{gisin04}.

\begin{figure}[htb]
\includegraphics[width=8cm]{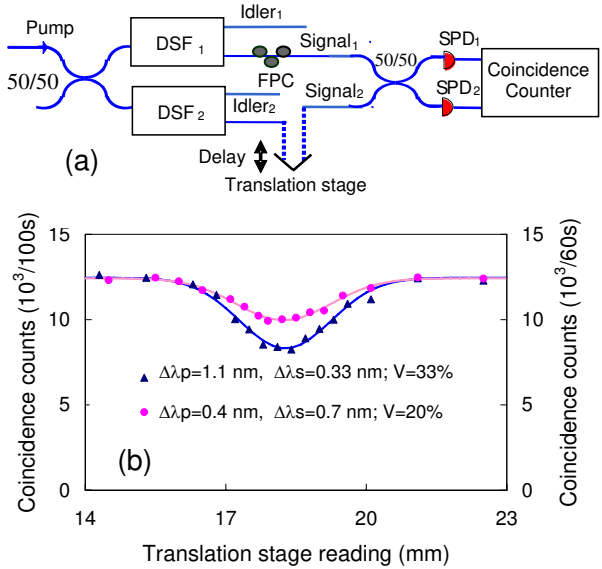}
\caption{(Color online) (a) Two-photon Hong-Ou-Mandel
interferometer with signal photons originated from two independent
DSF-based sources. (b) Twofold coincidences measured as a function of the
position of the translation stage. The error bar of the data is about the same as the size of the data points} \label{HOM-fig4}
\end{figure}

Finally, we perform a two-photon
HOM interference experiment with signal photons originated from two
independent DSFs (300\,m) to demonstrate that our source has the potential
application for QIP. With the experimental setup shown in
Fig.~\ref{HOM-fig4}(a), we measure two-fold coincidences in the outputs of the HOM interferometer versus
the position of the translation stage for two settings of the
signal and pump bandwidths, corresponding to $g^{(2)}=2$ and
$g^{(2)}=1.54$ in Fig.\ref{3}. Fig.\ref{HOM-fig4}(b) shows the HOM interference
result with visibility of $33\%$ and
$20\%$, respectively. The result is consistent with the theory of
fourth-order interference between two thermal field~\cite{Ou99}.

In conclusion, we have demonstrated a fiber source of photon pairs
with high brightness at telecom band. The normalized intensity
correlation function $g^{(2)}=2$ for individual signal (idler)
field is obtained, which indicates that the fields are in a single
temporal and spatial mode. With this fiber source, the ideal
visibility in two-photon HOM interference from independent sources
is achieved.

Because of lack of detectors, we cannot perform the HOM
interference with independent signal photons gated on the
detection of the idler photons, which ideally will give 100\%
visibility~\cite{Fulconis07}. Work is underway. Nevertheless, our two-photon experiment leads the same conclusion about the coherence property of the source as the four-photon experiment~\cite{Fulconis07}. Although the production rate and detection rate of
the photon pairs in current experiment is low, it can be
dramatically increased by using state-of-the-art pulsed fiber
lasers with a repetition rate over 10\,GHz and by using the newly
developed super-conducting single-photon detectors~\cite{Nam07}
with a higher detection efficiency.

This work was supported in part by NECT-060238, the State Key
Development Program for Basic Research of China (No. 2003CB314904),
the NSF of China (No. 60578024, No 10774111), and 111 Project B0704.
$^{*}$ X. Li's e-mail address is xiaoyingli@tju.edu.cn;
$^{\ddagger}$ Z. Y. Ou's e-mail address is zou@iupui.edu

\bibliographystyle{osajnl}

\bigskip

\end{document}